\documentclass[12pt,preprint]{aastex}

\newcommand{\jcd}{Christensen-Dalsgaard}

\shorttitle{Helioseismic constraints on solar abundances}
\shortauthors{Basu and Antia}

\begin{document}

\title{Constraining solar abundances using helioseismology}

\author{Sarbani Basu}
\affil{Astronomy Department, Yale University, P. O. Box 208101,
New Haven CT 06520-8101, U.S.A.}
\email{basu@astro.yale.edu}

\and

\author{H. M. Antia}
\affil{Tata Institute of Fundamental Research, Homi Bhabha Road,
Mumbai 400005, India}
\email{antia@tifr.res.in}

\begin{abstract}
Recent analyses of solar photospheric abundances suggest that the oxygen
abundance in the solar atmosphere needs to be revised downwards. In this
study we investigate the consequence of this revision on helioseismic
analyses of the depth of the solar convection zone and the helium abundance
in the solar envelope and find no significant effect. 
We also find that the revised abundances along with
the current OPAL opacity tables are
not consistent with seismic data. 
A significant upward revision of the opacity tables is required
to make solar models with lower oxygen abundance consistent with seismic
observations.

\end{abstract}

\keywords{Sun: abundances --- Sun: oscillations --- Sun: interior}

\section{Introduction}

Recent analyses of spectroscopic data using modern atmospheric models
have suggested that the solar abundance of oxygen and other abundant elements
needs to be revised downwards (Allende Prieto, Lambert \& Asplund 2001, 2002; 
Asplund et al.~2004). Asplund et al.~(2004) claim that the oxygen abundance should
be reduced by a factor of about 1.48 from the earlier estimates of
Grevesse \& Sauval (1998). The abundances of C, N, Ne
and Ar are also reduced.
The measured ratio of oxygen to hydrogen abundance, $[O/H]=8.66\pm0.05$,
is different from earlier estimates at approximately $3\sigma$ level.
As a result, the ratio (by mass) of heavy element to 
hydrogen abundance, $Z/X$, reduces from 0.023 to 0.0171, which causes
the heavy element abundance
in the solar envelope to reduce from $Z=0.017$ to 0.0126.
This will cause the opacity
of solar material to decrease, in turn reducing the depth of the
convection zone (henceforth CZ) in solar models. Bahcall \& Pinsonneault (2004) have
constructed a solar model using these revised abundances to find that
the depth of the CZ is indeed  reduced significantly as the CZ base
is found to be at $r_b=0.726R_\odot$, which is
inconsistent with the seismic estimate of $(0.713\pm0.001)R_\odot$
(Basu \& Antia 1997, henceforth BA97; Basu 1998). Asplund et al.~(2004) have argued
that their estimate refers to the abundances near the surface, while
because of diffusion the abundance could be much higher in the radiative
interior. 
While this may be true, it should be
noted that since the CZ is mixed on a rather small time-scale,
the abundances at the surface and above  the base of the CZ should 
be the same.  It is these abundances, coupled with the corresponding
opacity tables, that determine the depth of the CZ in a
solar model.
It may be possible to increase the hydrogen abundance
in the convection zone by increasing diffusion, thereby increasing the
convection zone depth, but in that case the helium content of
convection zone may reduce below the seismically estimated value.

Since the position of the CZ base, $r_b$,
has been determined very accurately through seismic analysis
(\jcd, Gough \& Thompson 1991; BA97) it provides a
constraint on the heavy element abundance or the opacities. In fact,
BA97 have shown that the then accepted solar abundances
along with the OPAL opacities (Iglesias \& Rogers 1996) are consistent with 
the helioseismic data. Thus it is of interest to test how
the reduction in abundance affects the conclusions. 
Before we do this, we examine whether a decrease in the abundances affects
the seismic estimate of the CZ depth or the estimate of
the helium abundance, $Y$.

\section{The technique}

To estimate the depth of the CZ, we use the technique
described by BA97. This technique requires the construction
of solar envelope models with prescribed values of $r_b$
and abundances along with
the known input physics. Solar envelope models are more useful than 
full solar models since envelope
models can be constructed with specific values of the CZ depth and $X$ (or $Y$)
and the CZ of the models  do not depend on other uncertainties like opacities,
treatment of diffusion, etc., in the radiative interior.
Since both the hydrogen abundance, $X$  and
the depth of the CZ can be determined seismically, we
can construct an envelope model with the seismic estimates of 
$r_b$ and $X$,
and then compare the sound speed and density profiles of the
model with the seismically obtained solar sound-speed and density
profiles to check for consistency.

The relative abundances of heavy elements are modified by the
reduction in the abundances of oxygen and other elements 
and we need to reconstruct the OPAL opacity tables for this new mixture of heavy elements 
(with $[O/H]=8.66$) which we refer to as MIX1.
In this mixture the
logarithmic abundances of C, N, O, Ne and Ar are reduced by 0.17
as compared to those given by Grevesse \& Sauval (1998), while abundances
of other elements are unchanged. 
It is possible that the abundances of
some of these elements change by a slightly different factor,
but that does not affect our conclusions.
We use the OPAL equation
of state (EOS) (Rogers \& Nayfonov 2002).
In principle, the EOS tables also need to be modified
in view of the change in mixture of heavy elements. We have not done
that  since the EOS is
not particularly sensitive to the detailed breakup of heavy element abundance.
Opacity, on the other hand, needs recalculation since in regions
of fully ionized H and He, the heavy elements are the predominant
contributors to opacity, though their contribution to the EOS is small.
Below the CZ we use the $X$
profile  determined from seismic data using the method of Antia \&
Chitre (1998) with the new heavy element abundances. We construct
solar envelope models with  CZ base position,
$r_b$, at $0.709R_\odot$, $0.711R_\odot$, $0.713R_\odot$,
$0.715R_\odot$ and $0.717R_\odot$ to serve as calibration models
for determining $r_b$. All these models have $Z/X=0.0171$ and
$X=0.74$ which is close to
the estimated value of hydrogen abundance. To determine
the helium abundance we use calibration models with $r_b=0.713R_\odot$
and $X=0.70,0.72,0.74,0.76,0.78$ and use the technique described
by Basu \& Antia (1995).

We also examine the consistency of similar models that have smaller reduction
(by 0.03) in the logarithmic abundances 
($[O/H]=8.80$ for these models). 
We refer to this mixture of heavy element abundances as MIX2.
The changes in abundances result in $Z/X=0.0218$ for these models.
We also  construct a few full solar models 
to check how they compare with seismic data.

For this work we use the observed frequencies  obtained by the
Global Oscillations Network Group (GONG) (Hill et al.~1996)  between
months 4 to 14, as well as frequencies obtained by the Michelson Doppler Imager (MDI)
from the first 360 days of its observation (Schou et al.~1998).

\section{Results}

Using the technique described by Basu \& Antia (1995) we first estimate
the helium abundance in the solar envelope using models with MIX1 abundance.
We find $X=0.7392\pm 0.0034$ using GONG data and $X=0.7385\pm 0.0034$ using
MDI data or  $Y=0.2482$ and 0.2489 respectively. The error-bars
include systematic errors including those caused by uncertainties in the
EOS (see Basu \& Antia 1995; Basu 1988 for details of uncertainties).
These results are consistent with earlier estimates that used models with completely
different heavy element abundances (D\"appen et al.~1991; Basu \& Antia
1995; Richard et al.~1996; Basu 1998).
We repeat the same exercise
using MIX2 abundances to find $X=0.7394$ (GONG) and 0.7386 (MDI). 
Thus it appears that the inferred hydrogen
abundance is insensitive to heavy element abundance of the calibration models, while the
helium abundance changes slightly due to change in $Z$.

To estimate the depth of the CZ
we use the technique described by BA97 that requires
the decomposition of the frequency differences between the calibration
models and the Sun in terms of two functions $H_1(w)$  and $H_2(\nu)$ that
depend on the sound-speed differences and surface structure respectively
(\jcd, Gough \& Thompson 1989). Here, $\nu$ is the frequency
and $w=\nu/(\ell+1/2)$, where  $\ell$ is the degree of the mode. 
The function $H_1(w)$ can be used to
determine the position of the CZ base.
Figure~1(a) shows the function $H_1(w)$ for
calibration models using MIX1 abundances. From this we can estimate
the position of the base of the CZ to be
$r_b=(0.7133\pm 0.0005)R_\odot$ (GONG) and $(0.7132\pm0.0005)R_\odot$ (MDI).
The error bars here include systematic errors as determined by Basu (1998).
These values are in good agreement with earlier
estimates and are evidently not affected by the revision of abundances.
The function $H_1(w)$ is a measure of the sound speed
differences between the models and the Sun, and Fig.~1(a) shows
that although the sound speed differences are small in the CZ,
there are significant systematic
differences  below the CZ. 
There is a small difference in
the sound speed even within the CZ.
This figure can be compared with Fig.~10 of BA97
to compare $H_1(w)$ for higher $Z$ models.
The difference is most likely due to reduction in opacities because
of reduced $Z$.

Although the seismic estimate of $X$  or CZ depth
 is not affected by the revision in abundances,
solar models constructed using the current abundances may not have
the correct depth of the CZ or the correct density profile
in the \hbox{CZ}. To check for this we 
compare the density profiles in these envelope models with that
obtained through seismic inversions, the results are shown in Fig.~1(b).
It is clear that the differences are very large.
The estimated error in density inversion
in most of the CZ is about 0.005.
In addition to this there could be systematic errors due to uncertainties
in $X$, $r_b$ and the EOS. An error of 0.0034 in $X$ causes
$\delta\rho/\rho\approx 0.012$,  an error of $0.0005R_\odot$ in
$r_b$ yields $\delta\rho/\rho\approx 0.005$, while uncertainties in
the EOS also give $\delta\rho/\rho\approx 0.005$.
Assuming that these errors are uncorrelated, we get a total error of
0.015 in $\delta\rho/\rho$.
Other uncertainties like the treatment of convection, turbulence and
the atmosphere etc., only affect the outermost layers
and hence are not included.
We have not included the effect of uncertainties in $Z$ and opacities as
we use seismic data to constrain these quantities by matching the density
profile (BA97) in a model with correct $X$ and $r_b$.
To get the correct density profile we need to increase the opacity
near the CZ base. We find that if the opacities
are increased by 19\%, then a MIX1 model with   $r_b=0.7133$ and $X=0.739$
has the correct density
profile (see Fig. 1b). Alternately, as in BA97 we can
estimate the value of $Z/X$ for the same relative abundances as MIX1
that will give the correct density
assuming that the OPAL opacities are correct
(i.e., we increase opacity by
increasing $Z$). We find that we need $Z/X=0.0214$ to get the
correct density profile (see  Fig.~1b).
We can also determine the  range of
opacities that give the density profile within acceptable limits
for each value of $Z/X$ and Fig.~3 shows the result.
Only
the $Z/X$ and opacity values in the shaded region are consistent
with seismic constraints. The point with error bars shows the
measured values assuming an uncertainty of 5\% in the opacity at
the CZ base.

We have repeated the analysis using models with MIX2 composition.
The results are shown in Fig.~2.
The position of the base of the CZ estimated using
GONG and MDI data is  $0.7135R_\odot$
and $0.7134R_\odot$ respectively. Thus
once again the  $r_b$ estimate is not significantly affected
by $Z$. From Fig.~2(b), it is clear that now we have better
agreement with  solar density in the \hbox{CZ}.
The function $H_1(w)$ is also essentially flat in the \hbox{CZ}.
However, the model with correct CZ depth
does not have the correct density.
In this case increasing opacity by 3.5\% or increasing $Z/X$ to 0.0228
makes it possible to get the correct density profile (see
Fig. 2b).
This value of $Z/X$ is close to
that found by Grevesse \& Sauval (1998).
The allowed region in the $Z/X$--opacity plane for MIX2 mixture is
also shown in Fig.~3 by the vertically shaded region.
The allowed region is just above that for MIX1 mixture. 

The above results were obtained with envelope models and it could be argued that
full solar models obtained from evolutionary calculations may give better results.
To check this we construct two 
full solar models, FULL1 and FULL2, with $Z/X$  and relative heavy element
abundance as in MIX1 and MIX2 respectively.
In these models, as with other standard solar models,
the surface $X$  and
the mixing-length parameter are adjusted to match the present day
solar radius and luminosity at an age of 4.6 Gyr.
These models were constructed with 
the Yale Rotating
Evolution Code in its non-rotating configuration (Guenther et al.~1992) and
includes diffusion of helium and heavy elements as per
Thoul, Bahcall \& Loeb (1994).
The sound-speed and density differences between these models and the Sun
are shown in Fig.~4.
The sound-speed differences  are dominated
by differences in the \hbox{CZ} depth. The
density differences are also affected by the differences in composition.
We can see that these models are worse than the envelope models.
Models FULL1 and FULL2 have
$r_b=0.7320R_\odot$ and $0.7217R_\odot$, and surface $X$ of 0.7505 and 0.7422 respectively.
We can see that
the model with the higher $Z/X$  is closer to the Sun, and we can argue that the
agreement improves with even higher oxygen abundance and  higher overall $Z/X$,  
which is in fact the case
(see, for example, model STD of
Basu et al.~2000, or model 20 of Winnick et al.~2002). Unlike models with higher 
$Z/X$, these models do
not reproduce the seismically determined value of $X$ in the solar envelope either.

Since Asplund et al.~(2004) have suggested that increased diffusion
may yield seismically consistent solar models with the  revised abundances,
we have constructed two models FULL1M and FULL2M with
diffusion coefficients increased by a factor of 1.65 and these models
are also shown in Fig.~4.
The positions of
the CZ bases are $0.7233R_\odot$ and $0.7138R_\odot$ for FULL1M and 
FULL2M respectively,
and the envelope $X$ is 0.7626 and 0.7519 respectively. Thus FULL2M has the
observed value of $r_b$, but the value of $X$ in the envelope is more
than $3\sigma$ beyond the seismically measured value.
It should be noted that we have increased the diffusion coefficients without any
physical justification.
We have also constructed static, full models of the Sun using
the seismically determined abundance profile (Antia \& Chitre 1998)
and these
are labeled INV1 and INV2 for MIX1 and MIX2 abundances.
These models have surface $X$ of 0.7680 and 0.7447 respectively and are also shown in Fig.~4.
Despite having the correct CZ depth,  the $X$ values in these models
are respectively, about $9\sigma$ and $2\sigma$ away from the seismically
inferred value.
Thus with the new abundances, we could not get the correct values of both $X$ and
CZ depth  in a  full solar model.
Models FULL1M, FULL2M, INV1, and INV2 are not standard solar
models since  the $X$ profile has been calculated using non-standard
procedures.

\section{Conclusions}

We have investigated the effects that the  revision of the  abundance of oxygen
and related elements in the solar atmosphere may have on seismic estimates
of the solar helium abundance and the depth of the \hbox{CZ}. We find
neither of these estimates to be sensitive to variations in $Z$.
We find that solar envelope models that have reduced abundances of oxygen 
and related elements do not have the correct density profile in the CZ despite
having the seismically determined CZ depth and surface $X$.
The density difference is about 10\%, which is more than 6 times
the estimated uncertainties in density.
In order to get a seismically consistent solar model it is necessary
to increase either the abundances of heavy elements or the
computed opacities for a given abundance ratio. 
Even for a much smaller reduction in the oxygen abundance ($[O/H]=8.80$, the MIX2 abundances)
it turns out that $Z/X$ needs to be increased
from 0.0218 to 0.0228 to match the sound speed and density
in the solar \hbox{CZ}. 
The region in the $Z/X$--opacity plane for both mixtures  that is consistent with 
seismic constraints 
is shown in Fig.~3.
The allowed region is not too sensitive
to variations in the oxygen abundance, but  to match the
seismic constraints either the opacity or the heavy element abundances,
or both need to be increased. From seismic constraints it
is not possible to decide between these possibilities. 

We find that the recent estimate of the value of
the oxygen abundance by Asplund et al.~(2004) along with the OPAL
opacities are not consistent with
seismic data. In fact, if current opacity tables are accepted,
no significant reduction in the oxygen abundance
from the values of Grevesse \& Sauval (1998) is  favored
by helioseismology.
Models constructed with the new abundances either have incorrect
CZ depth, or incorrect $X$, or incorrect density profile in the \hbox{CZ}.
Increasing the diffusion of heavy elements below the CZ base
to compensate for a reduction in the atmospheric abundance does not work,
the models fail to satisfy the $X$ constraint in the
solar envelope, even though they may have the correct CZ depth.
If the new values of abundances are confirmed, then 
opacity sources in the Sun will need to be re-investigated.  The intrinsic opacities
will need to be increased to counteract the decrease in opacity due to 
the reduction in heavy element abundances.
It could be argued that the abundances
in the atmosphere where the spectral lines are formed is different
from those in the CZ, due to some fractionation. This could resolve the problem
we face now. However we will be faced with the more fundamental problem 
of having no way of determining the CZ abundances of the Sun.

\acknowledgments

This work  utilizes data obtained by the Global Oscillation
Network Group project, managed by the National Solar Observatory
which is
operated by AURA, Inc. under a cooperative agreement with the
NSF.
This work also utilizes data from the Solar Oscillations
Investigation/ Michelson Doppler Imager (SOI/MDI) on the Solar
and Heliospheric Observatory (SOHO).  SOHO is a project of
international cooperation between ESA and NASA.
MDI is supported by NASA grants NAG5-8878 and NAG5-10483
to Stanford University.
We thank the OPAL group for the  opacity tables for the
different heavy element mixtures.
 This work was supported by
NSF grant ATM 0206130 to SB.

\clearpage

\begin{figure}
\plotone{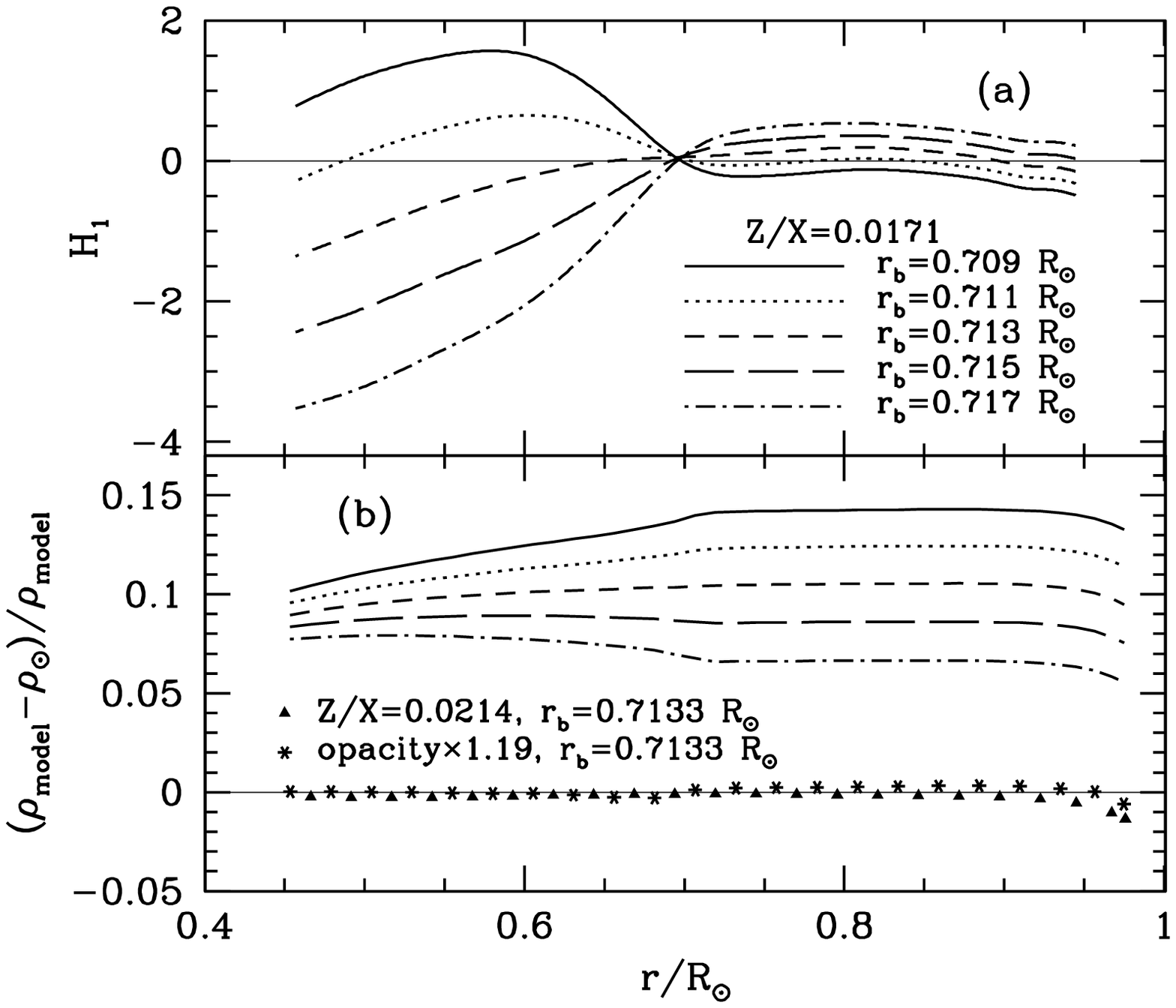}
\caption{Panel (a) --- The  function $H_1(w)$ obtained from frequency differences
between envelope models with MIX1 composition and the MDI
frequencies plotted as a function of lower turning point of the acoustic
modes. Panel (b) --- The relative difference in density between the  solar
envelope models with MIX1 composition and the  Sun. The different line styles 
are the same as in Panel(a).
The panel also shows the density differences for two other models,
one with $Z/X=0.0214$ and the other with opacities increased by 19\%.
\label{fig1}}
\end{figure}

\clearpage
\begin{figure}
\plotone{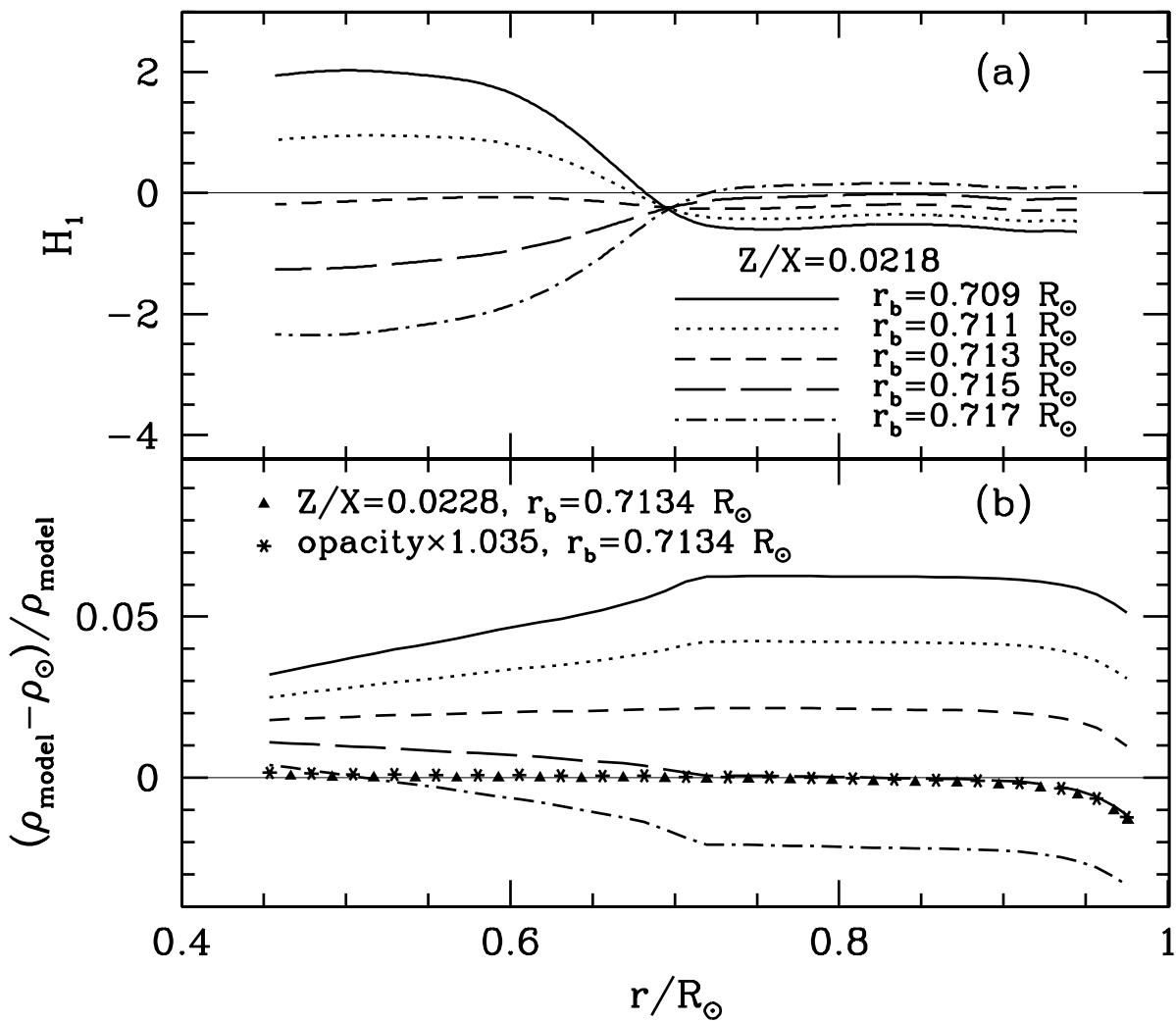}
\caption{
The same as Fig.~1, but for models with MIX2 composition.
The lower panel also shows the density differences for two other models,
one with $Z/X=0.0228$ and another with opacities increased by 3.5\%.
\label{fig2}}
\end{figure}

\clearpage
\begin{figure}
\plotone{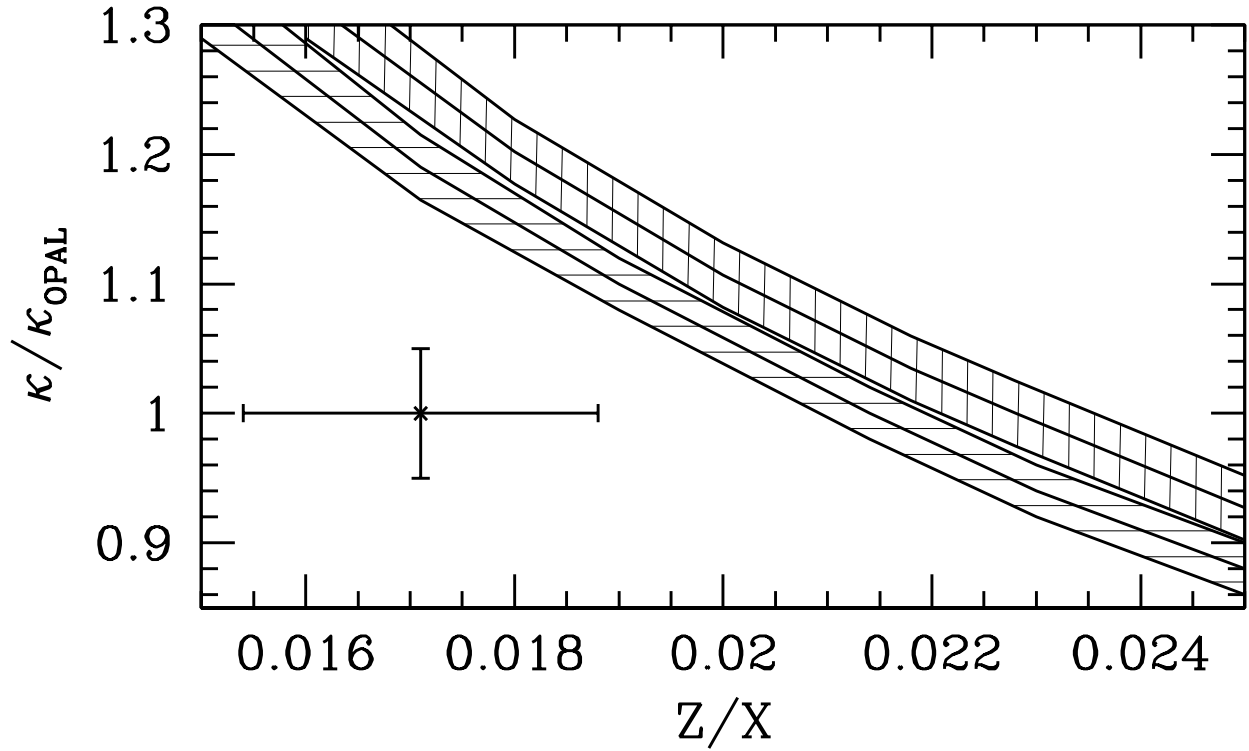}
\caption{The shaded area shows the allowed region in $Z/X$--opacity
plane that is consistent with seismic constraints. The horizontal
and vertical shadings, shows the region for MIX1 and MIX2 mixtures
respectively. The point with error bars shows current  values of
opacity and abundances.
\label{fig3}}
\end{figure}

\clearpage
\begin{figure}
\plotone{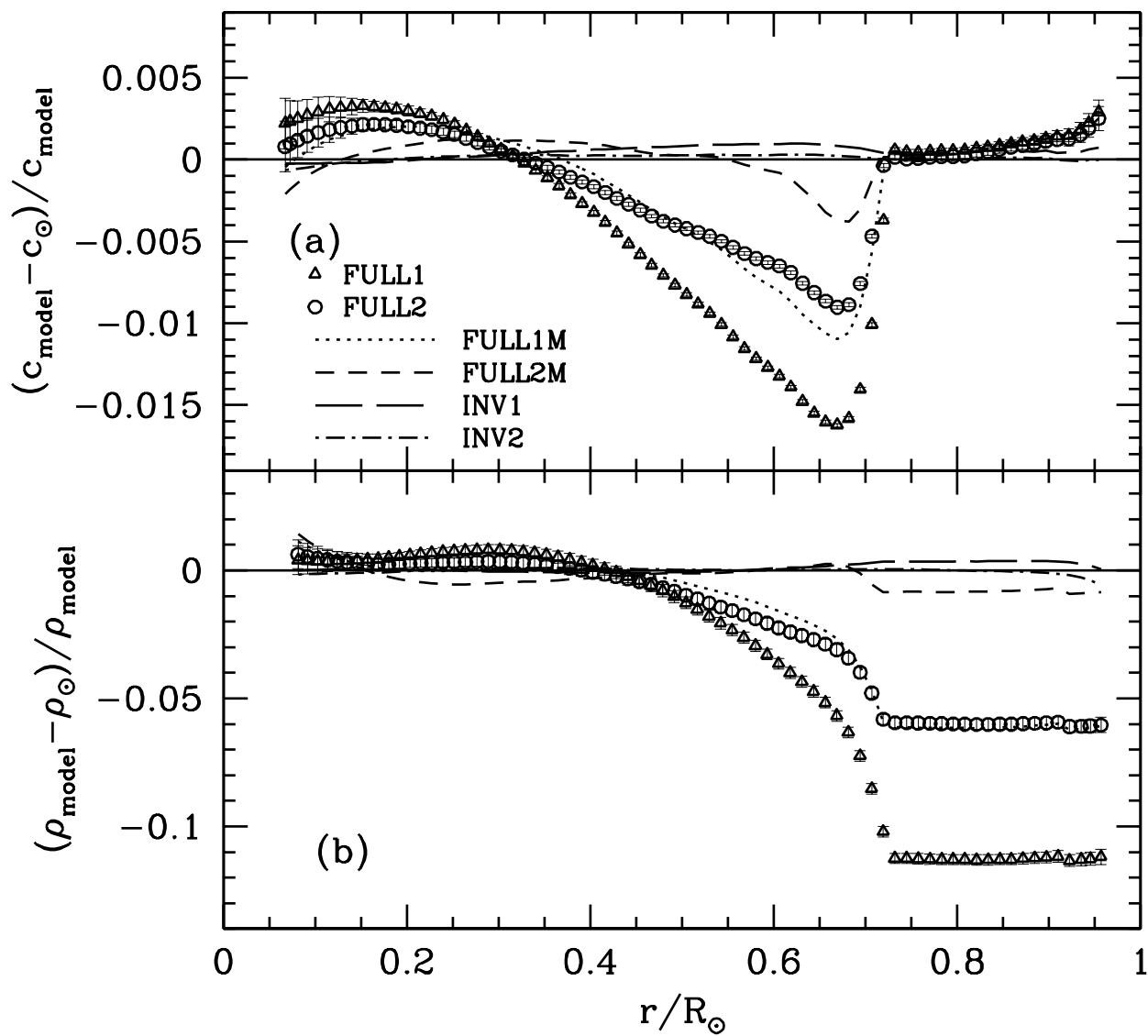}
\caption{ The  relative sound-speed and density differences between different
full solar models and the Sun. The points represent standard solar models, while the
lines show non-standard ones. Models FULL1, FULL1M and INV1 have 
 MIX1 composition and FULL2, FULL2M and INV2 have MIX2 composition.
\label{fig4}}
\end{figure}

\end{document}